\documentclass[pre,twocolumn,preprintnumbers,amsmath,amssymb,showpacs,nofootinbib,floatfix]{revtex4}

\usepackage{graphicx,bm}

\makeatletter
\def\graphicscale{\twocolumn@sw{0.3}{0.4}}
\def\graphicthreescale{\twocolumn@sw{0.3}{0.4}}

\begin{document}

\title{Finite-size scaling at the first-order quantum transitions of
quantum Potts chains}

\author{Massimo Campostrini,$^1$ Jacopo Nespolo,$^1$ 
Andrea Pelissetto,$^2$ and Ettore Vicari$^1$} 

\address{$^1$ Dipartimento di Fisica dell'Universit\`a di Pisa
        and INFN, Largo Pontecorvo 3, I-56127 Pisa, Italy}
\address{$^2$ Dipartimento di Fisica di ``Sapienza,'' Universit\`a di Roma
        and INFN, Sezione di Roma I, I-00185 Roma, Italy}

\date{\today}

\begin{abstract}
We investigate finite-size effects in quantum systems at first-order
quantum transitions.  For this purpose we consider the one-dimensional
$q$-state Potts models which undergo a first-order quantum transition
for any $q>4$, separating the quantum disordered and ordered phases
with a discontinuity in the energy density of the ground state.  The
low-energy properties around the transition show finite-size scaling,
described by general scaling ansatzes with respect to appropriate
scaling variables.  The size dependence of the scaling variables
presents a particular sensitiveness to boundary conditions, which may
be considered as a peculiar feature of first-order quantum
transitions.
\end{abstract}

\pacs{05.30.Rt,64.60.an,64.60.fd,64.60.De}

\maketitle



\section{Introduction}
\label{intro}

Zero-temperature quantum phase
transitions~\cite{Sachdev-book,Vojta-03} arise in many-body systems
with competing ground states controlled by nonthermal model
parameters.  They are of first order when the infinite-volume
ground-state properties are discontinuous across the transition point.
First-order quantum transitions (FOQTs) are of great interest, as they
occur in a large number of quantum many-body systems, such as quantum
Hall samples~\cite{PPBWJ-99}, itinerant ferromagnets~\cite{VBKN-99},
heavy fermion metals~\cite{UPH-04,Pfleiderer-05,KRLF-09}, etc.

Although FOQTs do not develop a diverging correlation length in the
infinite-volume limit, they show finite-size scaling (FSS) behaviors
around the transition point, which turn out to be particularly
sensitive to the boundary conditions~\cite{CNPV-14}.  Indeed, unlike
FSS at continuous transitions, the size dependence of the FOQT scaling
variables may significantly change when varying the boundary
conditions.  For example, in the case of the FOQTs of Ising chains,
driven by a magnetic field in their quantum ferromagnetic phase, we
have an exponential size dependence for open and periodic boundary
conditions, while it is power law for antiperiodic or kink-like
boundary conditions.  Actually, this particular sensitiveness to the
boundary conditions may be considered as a peculiar feature of FOQTs,
which qualitatively distinguishes their FSS behaviors from those at
continuous quantum transitions~\cite{SGCS-97,CPV-14}.  An
understanding of these finite-size properties is important for a
correct interpretation of experimental or numerical data when quantum
phase transitions are investigated in relatively small systems.

In Ref.~\cite{CNPV-14} we put forward general ansatzes for the FSS at
a FOQT driven by a generic parameter $g$.  Around the FOQT point
$g_c$, the relevant {\em scaling} variable $\kappa$ is expected to be
the ratio between the energy contribution $E_L$ of the perturbation
driving the transition [generally $E_L \approx (g-g_c) L^\zeta$ where
  $\zeta$ is an appropriate exponent] and the energy difference
({\em gap}) of the lowest states $\Delta_L\equiv E_1-E_0$ at
$g=g_c$. Then, around $g_c$, the gap satisfies the FSS ansatz
\begin{eqnarray}
\Delta(L,g) \equiv E_1(L,g) - E_0(L,g) \approx 
\Delta_{L} f_{\Delta}(\kappa),
\label{deltah}
\end{eqnarray}
where $\Delta_L\equiv \Delta(L,g_c)$, thus $f_\Delta(0)=1$.  Analogous
scaling behaviors apply to other quantities. For example, the
magnetization is expected to asymptotically behave as
\begin{eqnarray}
m(L,g) \approx m_0 f_m(\kappa),
\label{m0h}
\end{eqnarray}
where $m_0$ is a normalization which we identify with the
magnetization obtained approaching the transition point $g\to g_c$
from the ordered phase, after the infinite-volume limit. The above FSS
ansatzes represent the simplest scaling behaviors compatible with the
FOQT discontinuities arising in the infinite-volume limit. The
ansatzes can also be extended to allow for the
temperature~\cite{CNPV-14}. The particular sensitiveness to the
boundary conditions essentially arises from the energy gap at $g_c$
entering the scaling variable $\kappa$, whose finite-size behavior
depends crucially on the boundary conditions considered. Hence, once
the scaling variables are expressed in terms of the parameter driving
the transition and the size $L$, the $L$ dependence of the FSS may
significantly change according to the chosen boundary conditions.  The
above scaling ansatzes have been checked, and confirmed, at the FOQT
of Ising chains in the ferromagnetic phase, driven by the (odd)
magnetic field coupled to the order-parameter spin
operator.~\cite{CNPV-14}

In this paper we further develop this issue, extending the FSS
analysis to FOQTs driven by (even) temperature-like parameters, with a
discontinuity in the infinite-volume energy density of the ground
state. For this purpose we consider the quantum $q$-state Potts
chains~\cite{SP-81,IS-83}, which undergo FOQTs for a sufficiently
large number of states, i.e.  $q>4$~\cite{Baxter-73,Wu-82}.  They
provide a theoretical laboratory to test the scaling arguments at
FOQTs in a controlled framework, due to the relative simplicity of the
model for which the transition point is exactly known.  We observe FSS
described by the general ansatzes (\ref{deltah}) and
(\ref{m0h}). Moreover, like FOQTs of the Ising chains~\cite{CNPV-14},
its asymptotic size dependence turns out to be particularly sensitive
to the choice of the boundary conditions, giving rise to size
dependences with different power laws.

The paper is organized as follows.  In Sec.~\ref{qpch} we present the
quantum $q$-state Potts models, and discuss the boundary conditions of
finite chains. In Sec.~\ref{sizedep} we introduce some relevant
observables, and discuss their behavior at the FOQTs for $q>4$.  In
Sec.~\ref{fsstr} we present a numerical analysis around the transition
point, which supports the general FSS ansatzes at FOQTs.  In
Sec.~\ref{ordphase} we discuss the FSS behaviors at the FOQTs driven
by {\em parallel} magnetic fields in the ordered ferromagnetic phase,
which resemble those for Ising chains.  Finally, we draw some
conclusions in Sec.~\ref{conclu}.  A few appendices report some
details on the duality property of the quantum Potts chains, and the
numerical methods used to study them, which are based on the
density-matrix renormalization-group (DMRG) techniques~\cite{Sch-05}.

\section{The quantum Potts chain}
\label{qpch}

\begin{figure}
\vskip5mm
 \includegraphics{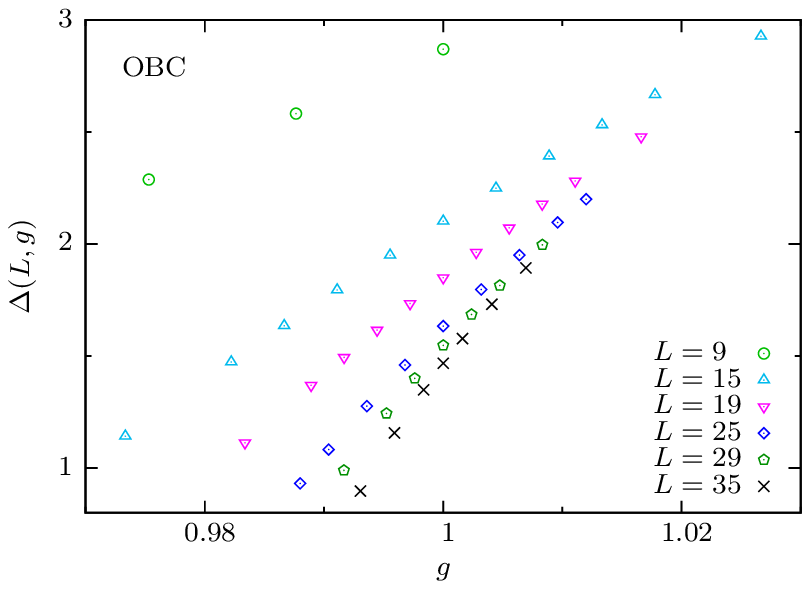}
\includegraphics{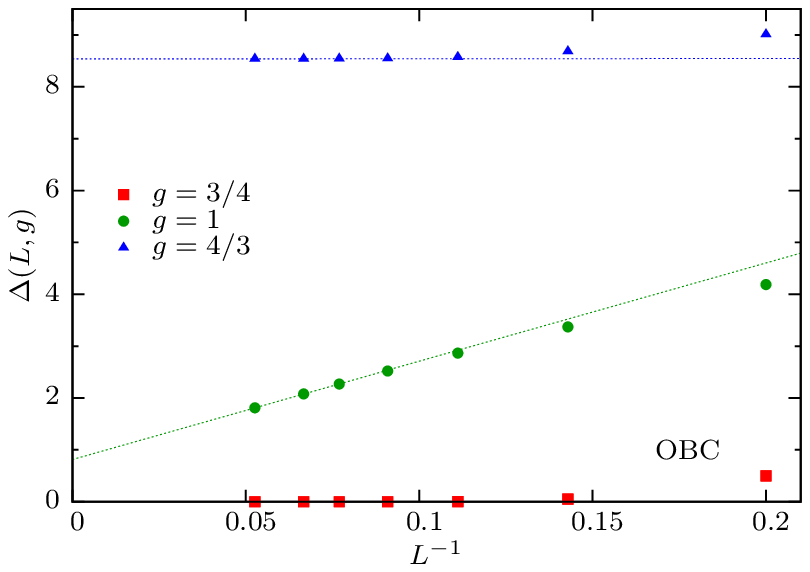} 
\caption{ (Top) The gap $\Delta(L,g)$ for the $q=10$ Potts chain with
  OBC, around $g=1$ and for several values of $L$.  (Bottom)
  $\Delta_L(L,g)$ vs $1/L$ at the transition point $g=1$, in the
  disordered ($g=4/3$) and ordered ($g=3/4$) phases.  The dotted lines
  are a linear fits of the data for the largest chains.  }
\label{gap-obc}
\end{figure}

\begin{figure}
\vskip5mm
\includegraphics{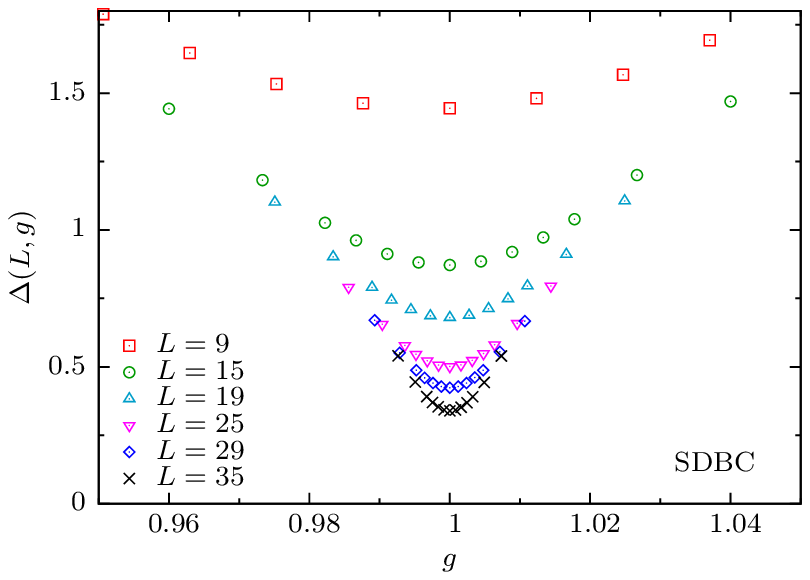} 
\includegraphics{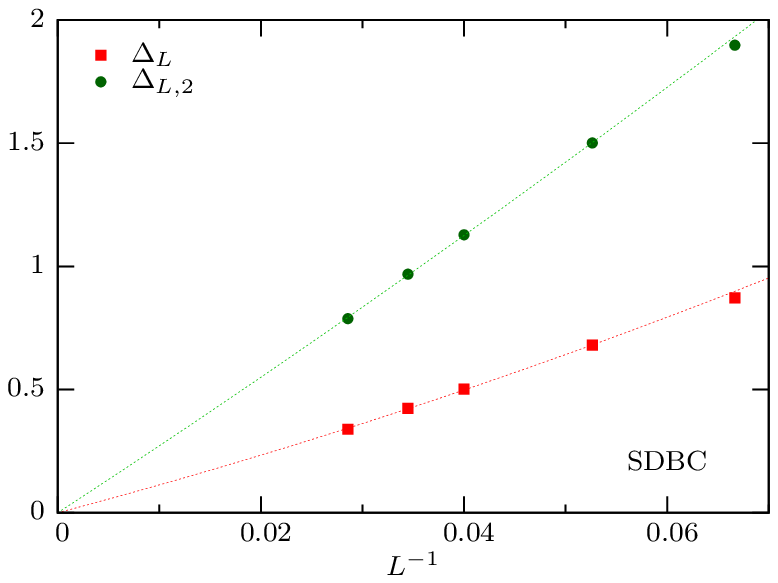} 
\caption{ Energy differences of the lowest states for the $q=10$ Potts
  chain with SDBC. The bottom figure shows data at $g=g_c$, for
  $\Delta_L \equiv E_1-E_0$ and $\Delta_{L,2}\equiv E_2-E_0$,
  providing evidence for an asymptotic $O(L^{-1})$ power-law behavior.
  The dotted lines are drawn to guide the eyes (they show fits to
  $p(L)=b_1/L + b_2/L^2$ of the data for $L>15$).  The top figure
  shows data for the gap $\Delta(L,g)$ around the transition point for
  several values of $L$.  }
\label{gap-dsbc}
\end{figure}

The quantum one-dimensional (1D) Potts chain is the quantum analog of
the classical 2D Potts model~\cite{Potts-52,Wu-82}
\begin{equation}
H_c = - J \sum_{\langle ij\rangle } \delta(s_i,s_j),
\label{cPotts}
\end{equation}
where the sum is over the nearest-neighbor sites of a square lattice,
$s_i$ are spin variables taking $q$ integer values, i.e.
$s_i=1,...,q$, and $\delta(m,n)=1$ if $m=n$ and zero otherwise.  A
corresponding 1D quantum Hamiltonian is derived from the {\em time}
continuum limit of the transfer matrix~\cite{SP-81}.  The quantum
$q$-state Potts chain involves $q$ states per site, which can be
labeled by an integer number $|n=1\rangle,...,|n=q\rangle$.  The
Hamiltonian of a chain of $L$ sites can be written
as~\cite{SP-81,IS-83}
\begin{equation}
H = - J_q  \sum_{j=1}^{L-1} \sum_{k=1}^{q-1} 
\Omega_j^k \Omega_{j+1}^{q-k} -
g \sum_{j=1}^L \sum_{k=1}^{q-1} M_j^k
\label{qPotts}
\end{equation}
where $J_q\equiv J/q$, $\Omega_j$ and $M_j$ are $q\times q$ matrices:
\begin{align}
 &\Omega = \delta_{m,n} \, \omega^{n-1},\qquad \omega = e^{i 2 \pi/q},
 \label{omega}\\
 &M = \delta_{m,{\rm Mod}(n-1,q)}=
\begin{pmatrix}
       0 & 1      &       &  \\
         & \ddots &\ddots  &   \\
         &       &        & 1 \\
       1 &        &        & 0 \\
      \end{pmatrix}.
\end{align}
These matrices commute on different sites and satisfy the
algebra:~\cite{SP-81} $\Omega_j^k \Omega_j^l = \Omega_j^{k+l}$, $M_j^k
M_j^l = M_j^{k+l}$, $ \Omega_j^q = M_j^q = \mathbb{I}$, $M_j^k
\Omega_j^l = \omega^{kl} \Omega_j^l M_i^k$.  For $q=2$ we recover the
quantum Ising chain.

Like Ising chains, the ground state of model (\ref{qPotts}) have two
phases: a disordered phase for sufficiently large values of $g$ and an
ordered phase for small $g$ where the system magnetizes along one of
the $q$ {\em directions}.  Indeed, for $g\to 0$ the ground state
(without further external fields) is degenerate, given by the $q$
states $\prod_{j} | n \rangle_j$, while for $g\to\infty$ it is
nondegenerate and given by $\prod_j \sum_{n=1}^{q} |n\rangle_j$.

Like Ising chains, quantum Potts chains satisfy a self-duality
property~\cite{SP-81}, i.e.
\begin{equation}
H(g/J_q) = (g/J_q) H(J_q/g),
\label{duality}
\end{equation}
see also App.~\ref{dualitysec}.  Therefore the transition point $g_c$
separating the two phases must be located at $g_c=J_q$.  The phase
transition is discontinuous (first order) for
$q>4$~\cite{Baxter-73,Wu-82}.  It becomes stronger and stronger with
increasing $q$, i.e.  the infinite-volume discontinuities increase
with increasing $q$.

The boundary conditions that we consider in our FSS study are specific
cases of the boundary Hamiltonian
\begin{equation}
H_b  = - J_q\left[ h_1 \sum_{k=1}^{q-1} \Omega_1^k 
+  h_L \sum_{k=1}^{q-1} \Omega_L^k\right] ,
\label{hb}
\end{equation}
which is added to the {\em bulk} Hamiltonian (\ref{qPotts}).  The
value $h_1=1$ (or $h_L=1$) is equivalent to having a further
fictitious site at $i=0$ (or $i=L+1$) with a fixed state
$|n=1\rangle$.  On the other hand, the value $h_1=0$ (or $h_L=0$) is
equivalent to having a further site $i=0$ (or $i=L+1$) with an
disordered unmagnetized state $\propto \sum_n |n\rangle$.  Therefore,
the case $h_1=h_L=1$ corresponds to fixed parallel boundary conditions
(FPBC) favoring the ordered phase, and the case $h_1=h_L=0$ to open
boundary conditions (OBC) favoring the disordered phase.

Self duality (\ref{duality}) is violated by finite chains with generic
boundary conditions, and in particular for generic values of
$h_1,\,h_L$. However, there is a notable exception, when $h_1=1$ and
$h_L=0$ (or viceversa), corresponding to mixed boundary conditions,
fixed and open at the two extremes (self-duality also requires a
spacial inversion $i\to L-i$), see App.~\ref{dualitysec}.  Thus the
duality relation (\ref{duality}), and in particular the relation for
the eigenspectrum
\begin{equation}
E_n(L,g/J_q) = (g/J_q) E_n(L,J_q/g),
\label{spduality}
\end{equation}
is satisfied for any finite chain when using these self-dual boundary
conditions (SDBC).

We define the magnetization of the ground state as
\begin{eqnarray}
&&m(L,g,x) = \langle {\cal M}_x \rangle,\label{mxdef}\\
&&{\cal M}_x= {q \delta(n_x,1) - 1\over q-1},
\quad \delta(n_x,1) = 
{1\over q} \sum_{k=1}^{q} \Omega_x^k,
\label{mxdefop}
\end{eqnarray}
such that $m =1$ if the site $x$ is along $n=1$, and $m =0$ if the
state at $x$ is an equally probable superposition of $q$ states. The
magnetization provides an order parameter: indeed it vanishes in the
disorder phase ($g>1$) and jumps discontinuously at the FOQT to a
nonzero value in the ordered phase, for $q>4$.  In particular,
\begin{equation}
m_0 = {\rm lim}_{g\to 1^-} \; {\rm lim}_{h\to 0}  \; {\rm lim}_{L\to\infty} \;
m(L,g,x)
\label{m0def}
\end{equation}
is non zero for $q>4$, where $h$ is a global magnetic field coupled to
the magnetization operator, e.g. described by the Hamiltonian term
$H_h = - h \sum_j \sum_{k=1}^{q} \Omega_j^k$.

The infinite-volume energy density of the ground state changes
discontinuously across the FOQT. We consider the definition
\begin{eqnarray}
&&e(L,g,x) = \langle {\cal E}_x \rangle,\label{energydens}\\
&&{\cal E}_x = \delta(n_x,n_{x+1}) = 
{1\over q}\sum_{k=1}^q \Omega_x^k \Omega_{x+1}^{q-k}.
\nonumber
\end{eqnarray}
The FOQT limits 
\begin{equation}
e_\pm = {\rm lim}_{g\to 1^\pm} \; {\rm lim}_{L\to\infty} \; e(L,g,x)
\label{epmdef}
\end{equation}
differ at the FOQTs of the Potts chains with $q>4$.  Their difference
$\Delta e\equiv e_+-e_-$ is the analog of the latent heat of
first-order classical transitions.

In the following we set $J_q=1$, thus the Hamiltonian depends on the
parameter $g$ only, and $g_c=1$.

\section{Size dependence at the transition point}
\label{sizedep}

In this section and next one we present a numerical analysis of the
$q=10$ Potts chains, employing DMRG techniques~\cite{Sch-05}.  Some
details on the implementation of DMRG methods for Potts chains are
reported in App.~\ref{dmrgapp}.

To begin with, we report the estimates of the infinite-volume
magnetization and energy values at the transition point,
cf. Eqs.~(\ref{m0def}) and (\ref{epmdef}), which characterize the
discontinuities of the ground state at the FOQT in the infinite-volume
limit. Infinite-size extrapolations of DMRG calculations at $g=1$ with
FPBC and OBC, favoring respectively the ordered and disordered phases,
provide the $q=10$ estimates
\begin{equation}
m_0 = 0.8572(1),\;e_- = 0.8060(1),\;e_+=0.3745(5).
\label{q10est}
\end{equation}

At the transition point $g_c=1$ for $q>4$, when the Potts chain
undergoes a FOQT, the energy difference ({\em gap}) of the lowest
states turns out to significantly depend on the choice of the boundary
conditions.

Let first consider OBC, which do not break the permutation symmetry of
the model.  Fig.~\ref{gap-obc} reports some numerical results for the
energy difference of the lowest states of the $q=10$ Potts chain.  The
data show that the energy difference of the lowest states approaches a
nonzero constant value in the disordered phase ($g>1$), i.e.
\begin{equation}
\Delta(L,g>1) = D(g) + O(L^{-1}).
\label{deltaobc}
\end{equation}
  This behavior is also observed at the transition point $g=1$, due to
  the fact that OBC favors disordered state.  Then, in the ordered
  phase ($g<1$) the ground state tends to be degenerate, and the gap
  gets exponentially suppressed, see also Sec~\ref{ordphase}. 
As shown in Fig.~\ref{gap-obc} the
  gap drops for $g\lesssim 1$, in a region of size $O(L^{-1})$ close
  to $g=1$, see also Ref.~\cite{IS-83}.  The gap at $g=1$ approaches a
  constant also in the case of FPBC, which favors the ordered phase.

\begin{figure}
 \includegraphics{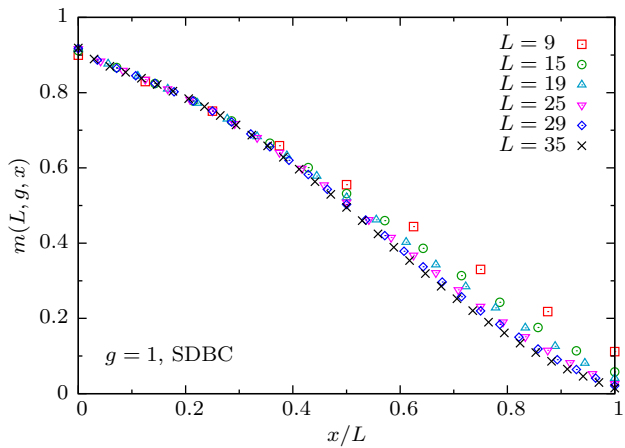}
 \caption{Space dependence of the magnetization (\ref{mxdef}) of the
   $q=10$ Potts chain with SDBC at the FOQT point $g=1$.  The data
   versus $x/L$ approach a scaling curve with increasing $L$.  }
\label{sdbcmagg1}
\end{figure}

A different behavior is found for SDBC. In this case the gap satisfies
the duality relation
\begin{equation}
\Delta(L,g) = g \Delta(L,1/g),
\label{dduality}
\end{equation}
which is a trivial consequence of Eq.~(\ref{spduality}).  Some results
for $q=10$ are shown in Fig.~\ref{gap-dsbc}.  The energy differences
of the lowest states at $g=g_c$ turn out to approximately decrease as
$1/L$ with increasing $L$, while they approach nonzero values for both
$g>1$ and $g<1$, in agreement with Eq.~(\ref{dduality}).  More
precisely, the data of $\Delta_L\equiv \Delta_{L,1}$ and
$\Delta_{L,2}$, related to the first and second excited state
respectively, fit the behavior
\begin{equation}
\Delta_{L,k} \equiv \Delta_k(L,g=g_c) = 
{c_k\over L} + {c_{k2}\over L^2} + ....
\label{dsbcgap}
\end{equation}
Although the data clearly distinguish the $O(L^{-1})$ asymptotic
behavior from other power laws, such as $O(L^0)$ or $O(L^{-2})$, see
Fig.~\ref{gap-dsbc}, they are not sufficient to obtain accurate
estimates of the coefficients $c_i$.  Some rough estimates are
$c_1\approx 9$ and $c_2\approx 3 c_1$ for $q=10$.

Analogous results are expected at the FOQTs for other values of
$q>4$. Once again, these results demonstrate how the size dependence
significantly changes when different boundary conditions are
considered at FOQTs.

We mention that some earlier analyses on the finite-size dependence at
the FOQTs of Potts chain appeared in Refs.~\cite{IS-83,IC-99},
focussing essentially on the large-$q$ limit.  Actually the
$O(L^{-1})$ finite-size behavior (\ref{dsbcgap}) of the gap appears to
contradict the large-$q$ results of Ref.~\cite{IC-99}, obtaining
$\Delta_L\sim 1/L^2$ for the same SDBC.  However, they were obtained
by first taking the $q\to\infty$ limit and then the large-$L$ limit.
This order of the limits does not necessarily reproduce the correct
results for the finite-size scaling at finite $q$, even at large $q$,
because the large-$q$ limit should be taken after the large-$L$ limit.

The magnetization (\ref{mxdef}) vanishes for OBC due to the $q$-state
permutation symmetry.  It is nonzero for SDBC which explicitly
(softly) breaks the $q$-state symmetry at one of the boundaries.
Fig.~\ref{sdbcmagg1} shows its space dependence at the transition
point of the $q=10$ Potts chain with SDBC: it goes from a number close
to one at the fixed $|n=1\rangle$ boundary to a number close to zero
at the opposite open extremity. Actually the data of the local
magnetization $m(L,1,L)$ at the open extremity appears to vanish,
apparently as $L^{-2}$.  Note that with increasing $L$ the data scale
as
\begin{equation}
m(L,g=1,x) \approx F(x/L).
\label{mg1xsca}
\end{equation}

\section{Finite-size scaling at the FOQT}
\label{fsstr}

First-order {\em classical} transitions, driven by thermal
fluctuations, show FSS behaviors analogous to those observed at
continuous
transitions~\cite{FB-72,Barber-83,Cardy-fss,Privman-90,PHA-91,PV-02},
with appropriate {\em critical exponents} related to the spatial
dimensionality $d$ of the system, see e.g.
Refs.~\cite{NN-75,FB-82,PF-83,FP-85,CLB-86,LK-91,BK-92,BNB-93,
  VRSB-93,CPPV-04,BDV-14}.  The extension of FSS to FOQTs is already
expected on the basis of the general quantum-to-classical mapping of
$d$-dimensional quantum systems into classical $(d+1)$-dimensional
systems with anisotropic slab-like geometry.  Indeed, a universal FSS
behavior emerges also at FOQTs~\cite{CNPV-14}.  However, unlike FSS at
continuous transitions, the depencence of the scaling variables on the
size $L$ may change significantly.  For example, at the FOQTs of Ising
chains driven by a parallel magnetic field in their ordered phase, the
scaling variables are related to the size $L$ by exponential or power
laws, depending on the properties of the low-energy spectrum for the
given boundary conditions.  In the case of the FOQTs of Ising chains,
this particular sensitiveness of the gap to the boundary conditions
was also noted in Refs.~\cite{CJ-87,LMMS-12}.

\begin{figure}
\vskip5mm
\includegraphics{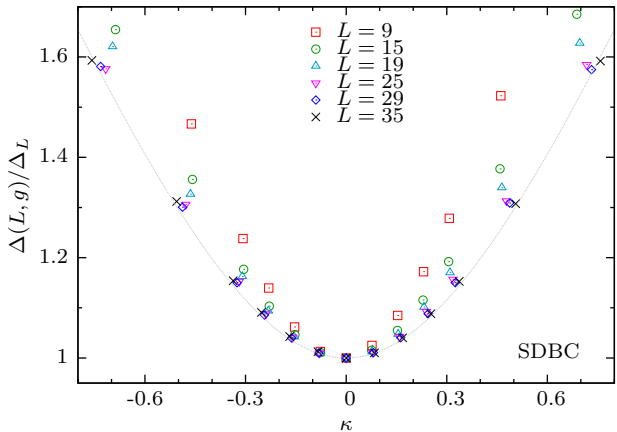}
\caption{FSS of the gap for the $q=10$ Potts chain with SDBC: we plot
  the ratio $\Delta(L,g)/\Delta_L$ versus $\kappa = (g-1)L/\Delta_L$
  for several values of $L$. They appear to approach an asymptotic FSS
  curve with increasing $L$.  The raw data were already shown in
  Fig.~\ref{gap-dsbc}.  The dotted line shows the optimal fit of the
  data of the largest lattice to the function $\sqrt{1+c\,\kappa^2}$
  with $c\approx 2.7$.  }
\label{fssgapsdbc}
\end{figure}

\begin{figure}
\includegraphics{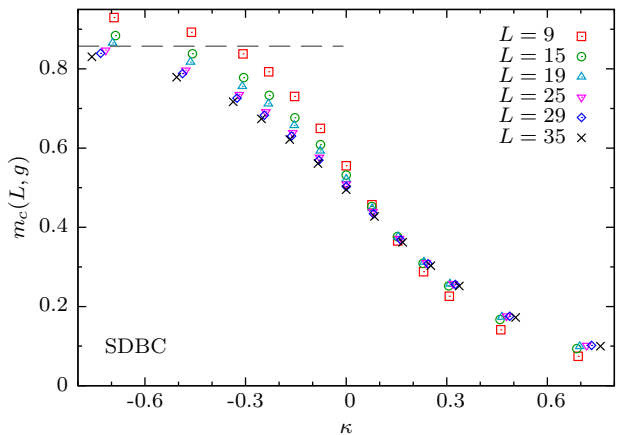}
\caption{ FSS of the central-site magnetization (\ref{mcdef}) for the
  $q=10$ Potts chain with SDBC.  The data are consistent with an
  asymptotic FSS as predicted by Eq.~(\ref{m0h}).  The dashed line
  indicates the inifinite-volume magnetization value,
  cf. Eq.~(\ref{q10est}).  }
\label{fssmagsdbc}
\end{figure}

\begin{figure}
\includegraphics{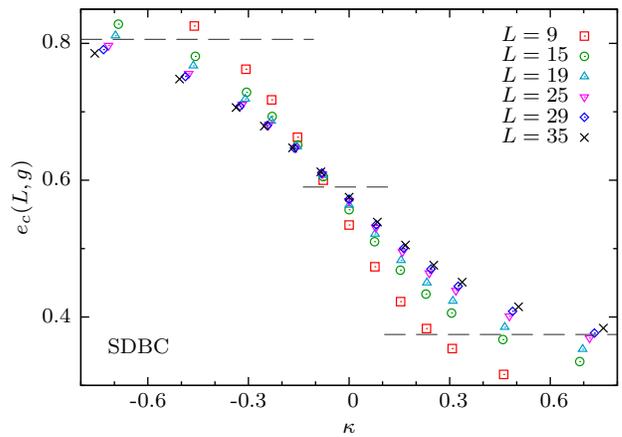}
\caption{ FSS of the energy density $e_c(L,g)$ at the center of the
  chain, cf. Eq.~(\ref{energydens}), for the $q=10$ Potts chain with
  SDBC.  The data approach a scaling function of $\kappa\equiv
  (g-1)L/\Delta_L$. They are consistent with the asymptotic behaviors
  given by the values $e_\pm$, cf.  Eq.~(\ref{q10est}), which are
  indicated by the dashed lines. At $g=g_c$, i.e. $\kappa=0$, the
  data approach the average value $(e_++e_-)/2\approx 0.590$ (dashed
  line around $\kappa=0$). }
\label{fssenesdbc}
\end{figure}

We now check the FSS scenario at the FOQTs of the $q=10$ Potts chain
by analyzing the numerical DMRG data of the energy differences of the
lowest states, the magnetization and energy density of the ground
state.

According to the general arguments outlined in Sec.~\ref{intro}, the
relevant scaling variable is expected to be
\begin{equation}
\kappa = {(g-1) L/\Delta_L},
\label{kappa}
\end{equation}
i.e. the ratio between $E_L\sim (g-1)L$, the energy contribution of
the perturbation driving the FOQT, and the gap $\Delta_L$ at the FOQT
point.

We first consider the model with SDBC, whose gap decays as $1/L$ at
the transition point.  Fig.~\ref{fssgapsdbc} shows that the data for
the ratio $\Delta(L,g)/\Delta_L$ approach a nontrivial scaling curve
in the large-$L$ limit when they are plotted versus the scaling
variable $\kappa$, nicely supporting the scaling ansatz
(\ref{deltah}).  Note that the scaling function $f_\Delta(\kappa)$
must be an even function of $\kappa$ due to the duality relation
(\ref{dduality}).  Since the disordered phase is generally gapped,
duality also implies that $f_\Delta(\kappa)\to \infty$ for
$\kappa\to\pm\infty$.  Actually the data for the largest lattices
suggest that
\begin{equation}
f_\Delta(\kappa) = \sqrt{1 + c \,\kappa^2},
\label{fdk}
\end{equation}
which is the simplest even function which diverges as $|\kappa|$,
matching the expected linear large-$L$ increase of the gap for finite
differences $g-1>0$.

The scaling behavior of the magnetization $m_c(L,g)$ at the center of
the chain,
\begin{equation}
m_c(L,g) \equiv m(L,g,x_c),
\label{mcdef}
\end{equation}
(with $x_c=(L+1)/2$ for chains with an odd number of sizes) is shown
In Fig.~\ref{fssmagsdbc}.  The plot of the data versus $\kappa$
definitely supports the scaling behavior (\ref{m0h}).  Asymptotically
for $\kappa\to -\infty$, the FSS curve appears to converge toward the
value $m_0$ defined in Eq.~(\ref{m0def}), thus $f_m(\kappa)\to 1$ for
$\kappa\to -\infty$ and $f_m(\kappa)\to 0$ for $\kappa\to\infty$.

Evidence of the FSS is also provided by the data of the energy density
$e_c(L,g)\equiv e(L,g,x_c)$ at the center of the chain,
cf. Eq.~(\ref{energydens}), see Fig.~\ref{fssenesdbc}. These data
appear to converge toward a scaling curve which has the values
$e_\pm$, cf. Eq.~(\ref{epmdef}), as asymptotics. Therefore, they
suggest the scaling behavior
\begin{equation}
e_c(L,g) \approx  f_e(\kappa)
\label{esca}
\end{equation}
with $f_e(\pm \infty)= e_\pm$ and $f(0)=(e_++e_-)/2$.

\begin{figure}
\includegraphics{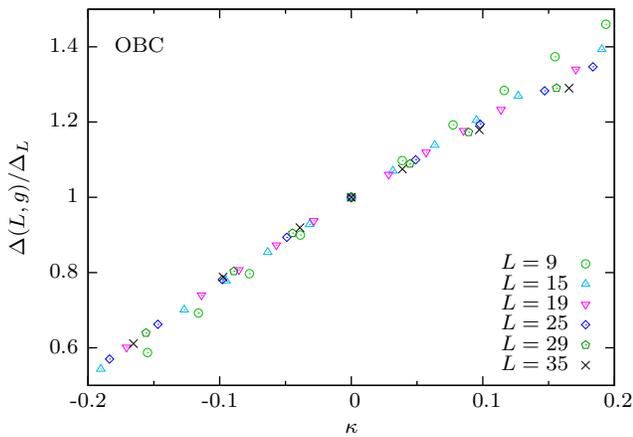}
 \caption{ FSS of the gap for the $q=10$ Potts chain with OBC.  The
   data approach a FSS curve with increasing the size $L$.  }
 \label{fssgapobc}
\end{figure}

Scaling behaviors are also observed in the case of OBC.  In this case
the magnetization vanishes by symmetry.  The results in
Fig.~\ref{fssgapobc} show a good evidence of the FSS ansatz
(\ref{deltah}) for the energy difference of the lowest
states. However, we stress that the actual size dependence of the FSS
is quite different from the case of SDBC where $\Delta_L = c/L +
O(L^{-2})$, because $\Delta_L = {\rm cost} + O(L^{-1})$ for OBC.  Also
the data of the energy density at the center of the chain show
evidence of scaling with respect to the variable $\kappa$, in
agreement with Eq.~(\ref{energydens}).

We expect that analogous results hold at the FOQTs for any $q>4$.
However, the values of $L$ where the asymptotic behavior begins being
observed may vary. We expect that larger sizes are necessary for
weaker FOQTs, i.e. smaller $q$, and smaller for stronger FOQTs,
i.e. larger $q$ values.

\section{FOQT driven by a magnetic field in the ferromagnetic phase}
\label{ordphase}

We now consider the FOQTs which occur in the ferromagnetic phase, i.e.
when $g<1$, driven by a {\em parallel  magnetic} field along one of the
$q$ directions, $n=1$ say.  In other words, we now consider a Potts
chain described by adding a magnetic term coupled to the global
operator $\sum_j \delta(n_j,1)$, i.e.
\begin{equation}
H_f = H - h \sum_{j=1}^L \sum_{k=1}^{q} \Omega_j^k
\label{hodef}
\end{equation}
where we use Eq.~(\ref{mxdefop}).  We consider OBC which do not break
the $q$-state permutation symmetry.  We study the interplay between
the finite size and the magnetic field $h$.
For any value of $q$, thus including the Ising chain for $q=2$, the
system has a line of FOQTs in the ordered ferromagnetic phase, i.e.
$g <1$, driven by the magnetic field $h$, which breaks the $q$-state
permutation symmetry, giving rise to a discontinuity of the magnetization
(\ref{mxdef}).  

The finite-size behavior at these FOQTs in the quantum Ising chain,
i.e.  the Hamiltonian (\ref{hodef}) for $q=2$, is analyzed in
Ref.~\cite{CNPV-14}, confirming the general FSS ansatzes
(\ref{deltah}) and (\ref{m0h}).  In the Ising case with boundary
conditions which do not break the Z$_{2}$ symmetry, such as OBC, the
two states with up and down global magnetization become degenerate for
$L\to \infty$.  In finite systems, an exponentially suppressed matrix
element between these states solve the degeneracy by mixing them,
giving rise to a phenomenon of avoided level crossings in many-body
systems. Thus the scaling functions can be determined exactly by
considering the theory restricted to the subspace spanned by the two
lowest-energy states.  Here we extend these computations to higher
values of $q$.

\begin{figure}
\includegraphics{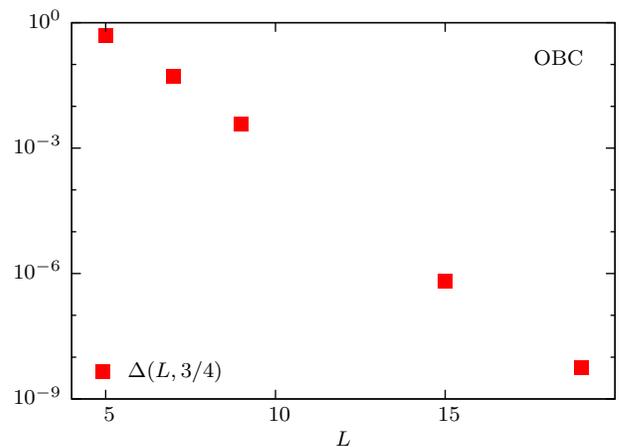}
\caption{ The gap of the $q=10$ Potts chain for $g=3/4$ and $h=0$, with OBC.
The data show an exponential suppression with increasing $L$.
}
 \label{gapg3o4}
\end{figure}
 
In the infinite-size limit without symmetry breaking terms, i.e. for
$h=0$, the low-energy energy spectrum of the ferromagnetic phase is
characterized by a degeneracy of the $q$ states $|\psi_n\rangle \equiv
\prod_j |n\rangle_j$ with $n=1,...,q$.  In a finite system, the
presence of a nonvanishing matrix element among these states,
essentially due to tunneling effects, lifts the degeneracy.  The
energy difference $\Delta_{L,k}$ of the lowest states for $h=0$
vanishes exponentially as $L$ increases, see, e.g.,  
Fig.~\ref{gapg3o4}, roughly as $\Delta_L \sim e^{-cL}$.
According to the general scaling arguments outlined in
Sec.~\ref{intro}, the scaling variable to describe the FSS at these
FOQTs is expected to be
\begin{equation}
\kappa \sim h L/\Delta_L.
\label{kappaqfe}
\end{equation}
Therefore the size dependence of $\kappa$ is exponential in this case,
i.e. $\kappa \sim h Le^{cL}$, analogously to the Ising $q=2$
case~\cite{CNPV-14}.

Since the energy differences $\Delta_{L,k}$ of the lowest $q-1$
excited states with the ground state are exponentially suppressed in
the large-$L$ limit, they are expected to be much smaller than the
energy differences between the higher excited states and the ground
state, i.e.
\begin{equation}
\Delta_{L,k\le q-1} \ll \Delta_{L,k\ge q}.
\label{approxde}
\end{equation}
Thus, we assume that, for sufficiently large $L$, the low-energy
properties in the crossover region around $h=0$, and for sufficiently
small $|h|$, are simply obtained by restricting the theory to the $q$
lowest-energy states $|n\rangle$.  The Hamiltonian restricted to this
subspace is a $q\times q$ matrix with the general form
\begin{eqnarray}
H_r = \left(
\begin{array}{l@{\ \ }l@{\ \ }l@{\ \ }l@{\ \ }l}
\varepsilon  - \beta  &  \quad -\delta   & \quad ...  &\quad -\delta \\
-\delta &  \quad\varepsilon + {\beta\over q-1}& \quad ... & \quad -\delta\\
...  &  \quad ... & \quad ...& \quad ... \\
-\delta &  \quad -\delta & \quad ... & \quad\varepsilon + {\beta\over q-1} \\
\end{array} \right) \; \label{hr}
\end{eqnarray}
where we take into account that the system must satisfy the $q$-state
permutation symmetry when $\beta=0$.  $\beta \sim h L$ represents the
perturbation induced by the magnetic field $h$.  $\delta\ge 0$ is a
small parameter which should vanish for $L\to \infty$ and $h=0$, in
order to obtain a degenerate ground state; its sign is such to have
the most symmetric state as ground state when $\beta=0$, as expected
in ferromagnetic models.  

\begin{figure}[tbp]
\includegraphics*[scale=\graphicscale]{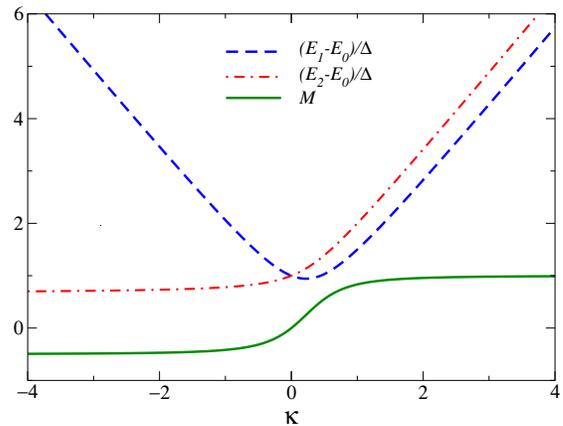}
 \caption{
The r.h.s. of Eqs.~(\ref{deemw1}) and (\ref{deemw2}),
and the scaling function of the magnetization (\ref{fsigma}).
}
\label{gapfe12}
\end{figure}

The gap can be easily obtained by diagonalizing $H_r$, taking the
differences of the eigevalues.  In particular, for $q=3$ and setting
$w = \beta/(3\delta)$, we obtain the energy differences
\begin{eqnarray}
&&{E_1 - E_0 \over 3 \delta}= \sqrt{ 1 - w  + {9\over 4} w^2},
\label{deemw1}\\
&&{E_2 - E_0\over   3 \delta}= {1\over 2} + {3\over 4} w+ {1\over 2}
\sqrt{ 1 - w  + {9\over 4} w^2},
\label{deemw2}
\end{eqnarray}
where $E_0$ is the lowest egenvalue.  These functions are shown in
Fig.~\ref{gapfe12}.  The above expressions agree with the scaling
ansatz (\ref{deltah}) provided that we identify
\begin{equation}
 3\delta \to \Delta_L,\qquad  w \to  \kappa.
\label{corrtbt}
\end{equation}
The scaling function $f_\Delta$ entering Eq.~(\ref{deltah}) can be
easily derived from the r.h.s. of Eqs.~(\ref{deemw1}) and
(\ref{deemw2}), taking their lowest value. Analogous expressions can
be straightforwardly obtained for higher $q$ values, although they
turn out to be more cumbersome.  

The ground-state magnetization can be obtained by computing the
expectation value of
\begin{equation}
{\cal M} = {q\delta(n,1)-1\over q-1}
\label{calm2}
\end{equation}
 on the lowest eigenstate.  In particular for
$q=3$ the scaling function in Eq.~(\ref{m0h}) reads
\begin{equation}
f_m(\kappa) = 
{\left( {3\over 2} \sqrt{ 1 - \kappa + {9\over 4}\kappa^2} - 
{1\over 2} + {9\over 4}\kappa\right)^2  - 1
\over 
\left( {3\over 2} \sqrt{ 1 - \kappa + {9\over 4}\kappa^2} - 
{1\over 2} + {9\over 4}\kappa\right)^2 + 2}
\label{fsigma}
\end{equation}
This function is shown in Fig.~\ref{gapfe12}.  The above FSS functions
are supposed to be asymptotically exact, i.e. for $L\to\infty$ keeping
$\kappa$ fixed. They are also expected to be universal with respect to
the values of the parameter $g<1$, which only enters through trivial
normalizations of the scaling variables and functions.  Scaling
corrections are expected to be suppressed in the FSS limit, by powers
of the inverse size.  Boundary effects in the case of OBC are expected
to be negligible for sufficiently large sizes.

The above $q$-level approximation provides also a framework to study
the unitary quantum dynamics when $h$ varies (in time) in a small
interval around $h = 0$.  Since only the $q$ lowest-energy states are
relevant, the dynamics is a straightforward extension of that
governing a two-level quantum mechanical system in which the energy
separation of the two levels is a function of time (Landau-Zener
effect ~\cite{LZeff}).

In conclusion the above analysis based on a $q$-level approximation,
restricting the Hamiltonian to the lowest almost-degenerate levels,
confirms the general features of the FSS scenario at FOQTs in
ferromagnetic phases driven by magnetic fields, generalizing the
results of Ref.~\onlinecite{CNPV-14} for the Ising chains
(corresponding to the $q=2$ case).

\section{Conclusions}
\label{conclu}

This paper continues the study of the finite-size scaling phenomena at
FOQTs. We extend the checks of the general FSS ansatzes put forward in
Ref.~\cite{CNPV-14}, considering FOQTs driven by an even
temperature-like parameter with a discontinuity in the ground-state
energy density.  For this purpose we consider the quantum $q$-state
Potts chains which undergo a FOQT when $q>4$, separating quantum
disordered and ordered phases. We present an extended numerical
analysis of the $q=10$ case, based on DMRG computations, considering
different boundary conditions: in particular OBC and SDBC, see
Sec.~\ref{qpch}.

Our analysis of the finite-size dependence of various low-energy
observables (such as the energy differences of the lowest states, the
magnetization and the energy density of the ground state) provide
evidence of asymptotic FSS behaviors for the various boundary
conditions considered, as predicted by the general arguments leading
to Eqs.~(\ref{deltah}), (\ref{m0h}), and (\ref{esca}).  However, this
fact does not prevent the system from having different functional
dependences on the finite size. Indeed, the actual size dependence is
essentially due to the size dependence of the scaling variable
$\kappa$, cf. Eq.~(\ref{kappa}), and in particular of the {\rm gap}
$\Delta_L$ entering its definition.  This leads to different power
laws when passing from open to mixed self-dual boundary conditions.
The general sensitiveness of the FSS at FOQTs may be considered as
their peculiar feature with respect to continuous quantum transitions.

We also discuss the FSS behavior at the FOQTs driven by parallel
magnetic fields in the ordered ferromagnetic phase, for any $q$, 
thus also for the Ising chain.  We argue that the FSS
behavior can be obtained by extending the two-level approximation used
in the case of Ising chains~\cite{CNPV-14}, to a $q$-level approximation. Like the
analogous FOQTs of Ising chains, we confirm the existence of FSS, with
a scaling variable which depends exponentially on the size of the
chain.

We stress that the FSS behavior at FOQTs is observed for relatively
small sizes. Indeed, chains of size up to a few tens turn out to be
sufficient to show the asymptotic FSS behavior.  Thus, even small
systems may show definite signatures of FOQTs, as also argued in
Refs.~\cite{CNPV-14,IZ-04,LMD-11}.  This makes FSS at FOQTs
particularly interesting, because it may require only modest-size
systems to be observed.  For this reason, the present results may be
relevant for quantum simulators, where controllable quantum systems,
usually of modest size, are engineered to reproduce particular
Hamiltonians~\cite{GAN-14,Islam-etal-11,Simon-etal-11,LMD-11}.
Furthermore, in quantum computing, some algorithms, notably the
adiabatic ones, rely on a sufficiently large
gap~\cite{Bloch-08,YKS-10,AC-09,LMMS-12}, and thus fail at FOQTs.  The
scaling theory that we present may help to understand how this occurs
in finite systems.

We finally mention that an interesting extension of this study
concerns the effects of inhomogeneous conditions at FOQTs, such as
those induced by space dependent external fields.  Nontrivial scaling
phenomena emerge also in this case, with respect to the length scale
of the induced inhomogeneity, in the space transition region between
the two phases~\cite{CNPV-15}.

\appendix

\section{Self duality of quantum Potts chains}
\label{dualitysec}

We first consider the finite-chain Potts model defined by the
Hamiltonian
\begin{equation}
H_{\rm sd} = - J_q \sum_{j=1}^{L-1} \sum_{k=1}^{q-1} 
\Omega_j^k \Omega_{j+1}^{q-k} -
g \sum_{j=1}^{L-1} \sum_{k=1}^{q-1} M_j^k
\label{qPottssd}
\end{equation}
Notice that the difference with the Hamiltonian (\ref{qPotts}) is just
relegated to the second sum which does not contain the lattice size
$i=L$.  In this case one can define an exact duality transformation,
which maintains the form of the Hamiltonian (\ref{qPottssd}),
interchanging the role of the two sums, thus $J_q \to g$ and $g\to
J_q$. This is achieved by defining the new operators
$\widetilde{M}_{l}$ and $\widetilde{\Omega}_l$,~\cite{SP-81} formally
located between two adjacent sites, such that
\begin{equation}
\widetilde{M}_{j+1/2}^k = \Omega_j^k\Omega_{j+1}^{q-k},\qquad
\widetilde{\Omega}_{j+1/2}^k = \prod_{i\le j} M_i^{q-k}, 
\label{dualtra}
\end{equation}
which satisfy the same algebra of the original operators $\Omega_j$
and $M_j$. The inverse dual transformation reads
\begin{equation}
M_{l+1/2}^k = \widetilde{\Omega}_{l}^k\widetilde{\Omega}_{l+1}^{q-k},\qquad
\Omega_{l-1/2}^k = \prod_{i\ge l} \widetilde{M}_i^{q-k}.
\label{dualtrain}
\end{equation}
Note that, in order to get an Hamiltonian formally identical to
$H_{\rm sd}$, one should also perform a space inversion $j\to L-j$.
This duality properties imply that a unitary transformation $U$ exists
such that
\begin{equation}
U H_{\rm sd}(J_q,g) U^\dagger  = H_{\rm sd}(g,J_q) = g H_{\rm sd}(1,J_q/g),
\label{uh1}
\end{equation}
which also implies the spectrum relation (\ref{spduality})
for $J_q=1$.

Since $[\Omega_L,H_{\rm sd}]=0$, the spectrum of $H_{\rm sd}$ is
degenerate, with a degeneracy associated with the $q$ eigenstates of
$\Omega_L$.  This implies that the Hilbert space can be decomposed
into $q$ subspaces ${\cal H}_{n}$, such that $\Omega_L \psi_n =
\omega^{n-1} \psi_n$, cf. Eq.~(\ref{omega}).  If we restrict $H_{\rm
  sd}$ to ${\cal H}_{n=1}$, we obtain the model described by the
Hamiltonian (\ref{qPotts}) supplemented with the boundary term
(\ref{hb}) with $h_1=1$ and $h_L=0$, for a chain of length $L-1$.
This is already sufficient to show that the finite Potts chain with
SDBC is self dual, thus its spectrum satisfies the duality relation
(\ref{spduality}).  Of course, this could have been demonstrated by
directly using the duality transformations (beside those given in
Eq.~(\ref{dualtra}) one should use $\widetilde{M}_L^k =
\Omega_L^{q-k}$).

\section{DMRG computations of the Potts chains}
\label{dmrgapp}

\begin{figure}
\vskip5mm
\includegraphics{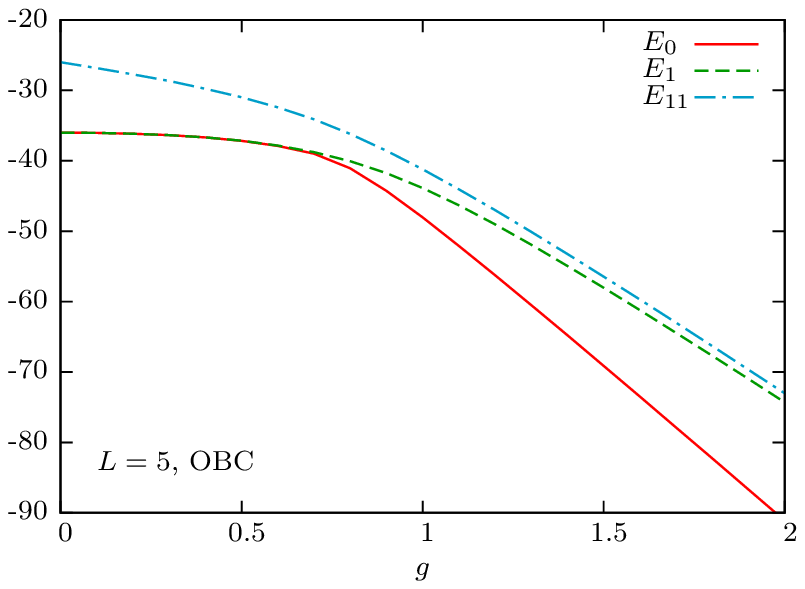} 
 \includegraphics{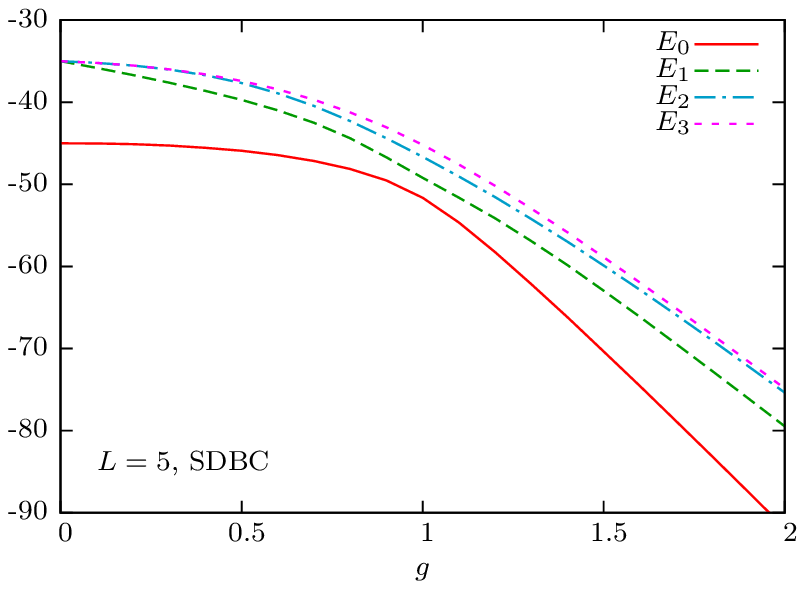}
\caption{ Energy levels of the $q$-state Potts chains with $L=5$, for
  OBC (Top) and SDBC (Bottom).  (Top) The non degenerate, even, ground
  state level $E_0$ approaches the $(q-1)$ times degenerate multiplet
  of energy $E_1$ in the ordered phase $g<1$. (Bottom) The ground
  state, together with the levels $E_1$ and $E_2$, remain non
  degenerate at any $g>0$. $E_3$ is the lowest multiplet in the SDBC
  spectrum.  }
\label{elevL5}
\end{figure}

We use a standard implementation of the DMRG algorithm~\cite{Sch-05},
adapted to the $q$-state Potts model.  The computation involves
solving eigenproblems of a size that rapidly grows with $q$ as $m^2
q^2$, where $m$ is the number of states used in the DMRG base.  We
used up to $m=200$ for our DMRG computation of the $q=10$ Potts
chains.  In the worst case, corresponding to the longest chains, 
the truncation error (i.e., the sum of the discarded
eigenvalues of the reduced density matrix) is $O(10^-9)$.

The energies $E_n$ of the excited states (and the differences
$\Delta_n(L,g)=E_n-E_0$ as a consequence) are the quantities affected
by larger convergence errors, whereas ground state properties, such as
the ground state energy, the energy densities and the magnetizations,
converge much more rapidly.  The systematic error on a DMRG
measurement can be estimated by looking at the progression of its
convergence with successive iterations of the algorithm and as $m$ is
gradually increased.  We benchmarked this heuristics by comparing DMRG
measurements with exact values on small chains, obtained with a
diagonalization code.

To give an idea of the computational effort and accuracy of our
numerical work, we mention that the DMRG computation for $L=35$ chain
with SDBC at $g_c=1$ required a few hundred hours of CPU time of a
standard processor to reach the truncation error of $2\times 10^ {-9}$
for $m=200$.  This allowed us to obtain the estimates $E_0 =
-346.484988(2)$ (where the number in parenthesis quantifies the
systematic error due to the truncation of the states, corresponding to
a relative precision of $0.6\times 10^{-8}$ for $E_0$), $E_1 =
-346.1459(7)$, $E_2 = -345.696(2)$ for the lowest energy levels, thus
$\Delta_L= 0.3390(7)$ and $\Delta_{L,2} =0.789(2)$, and $m_c =
0.529723(3)$, $e_c = 0.593785(1)$ for the magnetization and energy
density at the center of the chain.  Note that this accuracy allows us
to determine the energy differences of the lowest states with a
relative precision of $O(10^{-3})$ for $\Delta_L=E_1-E_0$ and
$O(10^{-2})$ for $\Delta_{L,2}=E_2-E_0$, which we deem sufficient for
the observation of the scaling properties of the Potts model.  We
stress that the calculation of the energy differences, i.e. the gap,
is essential to check the FSS ansatzes, thus we could not restrict our
computations to the ground-state expectation values, which are usually
much more precise.

Since the precision turns out to rapidly degrade with increasing
$L$, making the computations very demanding to reach a satisfactory
accuracy, we restricted our $q=10$ study to chain up $L=35$ sites.
Fortunately, these relatively small sizes turn out to be already
sufficient to provide a clear evidence of FSS at the FOQTs.

The low energy spectrum of the Potts chain with OBC is characterised
by highly degenerate multiplets~\cite{IS-83} (see also
Fig.\ \ref{elevL5}), which make DMRG computations particularly
difficult.  This is shown in Fig.\ \ref{elevL5}, where we show the
energy of the lowest levels for OBC and SDBC with $L=5$.  Analogous
behaviors apply for any finite $L$.  The DMRG algorithm makes
extensive use of sparse eigensolvers, whose rate of convergence and
numerical stability depend on the spacing between the different
eigenvalues.  The system with SDBC is not affected by this issue
thanks to the fixed spin at one end of the chain, which lifts the
degeneracy.  To give an idea of this effect, we performed a test run
on a Potts chain with $q=10, L=9$ and $g=0.98$ with both SDBC and OBC,
completing two sweeps with $m=20$ basis states and keeping all other
simulation parameters unchanged.  For $L\to\infty$ and $g<1$, the OBC
system is expected to exhibit a $q$-times degenerate ground state.
However, finite size corrections make the ground state non degenerate,
with a $(q-1)$-times degenerate first excited level, as shown in
Fig.~\ref{elevL5}.  The run with SDBC takes approximately 180 seconds
to complete, while the OBC run takes almost four times as long on the
same machine.  This efficiency gap is expected to widen for larger
values of $m$, as the computation progresses.  This problem can be
partly mitigated by changing the internal parameters of the
eigensolver routine, at the expense of some extra memory usage.  For
these reasons we obtain less precise results for the OBC.  For
example, for $L=35$ at $g=1$ we obtain $E_0=-342.1121(6)$ and
$E_1=-340.645(1)$.


\begin{thebibliography}{99}

\bibitem{Sachdev-book} 
S. Sachdev, {\em Quantum Phase Transitions}, (Cambridge University
Press. 2011, 2nd ed.)

\bibitem{Vojta-03}
M. Vojta, Rep. Prog. Phys. {\bf 66}, 2069 (2003).

\bibitem{PPBWJ-99} V. Piazza, V. Pellegrini, F. Beltram,
  W. Wegscheider, T. Jungwirth, and A.H. MacDonald, Nature {\bf 402},
  638 (1999).

\bibitem{VBKN-99}
T. Vojta, D. Belitz, T.R. Kirkpatrick, and R. Narayanan,
Ann. Phys. (Leipzig) {\bf 8}, 593 (1999). 

\bibitem{UPH-04}
M. Uhlarz, C. Pfleiderer, and S.M. Hayden,
Phys. Rev. Lett. {\bf 93}, 256404 (2004).

\bibitem{Pfleiderer-05}
C. Pfleiderer, J. Phys.: Cond. Matter
{\bf 17}, S987 (2005). 

\bibitem{KRLF-09}
W. Knafo, S. Raymond, P. Lejay, and J. Flouquet,
Nature Phys. {\bf 5}, 753 (2009).

\bibitem{CNPV-14}
M. Campostrini, J. Nespolo, A. Pelissetto, and E. Vicari,
Phys. Rev. Lett. {\bf 113}, 070402 (2014).

\bibitem{SGCS-97}
S.L. Sondhi, S.M. Girvin, J.P. Carini, and D. Shahar,
Rev. Mod. Phys. {\bf 69}, 315 (1997).

\bibitem{CPV-14} M. Campostrini, A. Pelissetto and E. Vicari,
  Phys. Rev. B {\bf 89}, 094516 (2014).

\bibitem{SP-81}
J. S\'olyom and P. Pfeuty, Phys. Rev. B {\bf 24}, 218 (1981).

\bibitem{IS-83}
F. Igl\'oi and J. S\'olyom, J. Phys. C: Solid State Phys.
{\bf 16}, 2833 (1983).

\bibitem{Baxter-73}
R.J. Baxter, J. Phys. C: Solid State Phys. {\bf 6}, L445 (1973);
R.J. Baxter, H.N.V. Temperley, and S.E. Ashley,
Proc. R. Soc. Lond. A {\bf 538}, 535 (1978).

\bibitem{Wu-82}
F.Y. Wu, Rev. Mod. Phys. {\bf 54}, 235 (1982).

\bibitem{Sch-05}
U. Schollw\"ock,  Rev. Mod. Phys. {\bf 77}, 259 (2005).

\bibitem{Potts-52}
R.B. Potts, Math. Proc. Camb. Phil. Soc. {\bf 48}, 106 (1952)

\bibitem{IC-99} F. Igl\'oi and E. Carlon, Phys. Rev. B {\bf 59}, 3783
  (1999).

\bibitem{FB-72}
M. E. Fisher and M. N. Barber, Phys. Rev. Lett. {\bf 28}, 1516 (1972).

\bibitem{Barber-83}
M. N. Barber, in {\em Phase Transitions and Critical Phenomena},
edited by C. Domb and J. L. Lebowitz (Academic Press, New York, 1983),
Vol. 8.

\bibitem{Cardy-fss}
{\em Finite Sise Scaling}, edited by J. Cardy,
(Elsevier, 1988).

\bibitem{Privman-90} V. Privman ed.,
{\em Finite Size Scaling and Numerical Simulation of Statistical Systems}
\/ (World Scientific, Singapore, 1990).

\bibitem{PHA-91}
V.~Privman, P.~C.~Hohenberg, and A.~Aharony,
in {\em Phase Transitions and Critical Phenomena}, Vol.\ 14,
edited by C.~Domb and J.~L.~Lebowitz (Academic Press, New York, 1991).


\bibitem{PV-02} A. Pelissetto and E. Vicari, Phys. Rep. {\bf 368}, 549 (2002).

\bibitem{NN-75}
B. Nienhuis and M. Nauenberg, Phys. Rev. Lett. {\bf 35}, 477 (1975).

\bibitem{FB-82}
M.E. Fisher and A.N. Berker,
Phys. Rev. B {\bf 26}, 2507 (1982);

\bibitem{PF-83}
V. Privman and M. E. Fisher, 
J. Stat. Phys. {\bf 33}, 385 (1983).

\bibitem{FP-85}
M. E. Fisher and V. Privman,
Phys. Rev. B {\bf 32}, 447 (1985).

\bibitem{CLB-86}
M.S.S. Challa, D.P. Landau, and K. Binder,
Phys. Rev. B {\bf 34}, 1841 (1986).

\bibitem{LK-91}
J. Lee and J.M. Kosterlitz,  Phys. Rev. B {\bf 43}, 3265 (1991).

\bibitem{BK-92}
C. Borgs and R. Koteck\'y,
Phys. Rev. Lett. {\bf 68}, 1734 (1992).

\bibitem{BNB-93}
A. Billoire, T. Neuhaus, and B.A. Berg,
Nucl. Phys. B {\bf 396}, 779 (1993).

\bibitem{VRSB-93}
K. Vollmayr, J.D. Reger, M. Scheucher, and K. Binder,
Z. Phys. B {\bf 91}, 113 (1993).


\bibitem{CPPV-04}
P. Calabrese, P. Parruccini, A. Pelissetto, and E. Vicari,
Phys. Rev. B {\bf 70}, 174439 (2004).

\bibitem{BDV-14}
C. Bonati, M. D'Elia, and E. Vicari,
 Phys. Rev. E {\bf 89}, 062132 (2014).

\bibitem{CJ-87}
G.G. Cabrera and R. Jullien, 
Phys. Rev. B {\bf 35}, 7062 (1987). 

\bibitem{LMMS-12}
C.R. Laumann, R. Moessner, A. Scardicchio, and S. L. Sondhi, 
Phys. Rev. Lett. {\bf 109}, 030502  (2012)

\bibitem{LZeff}
C. Zener, Proc. R. Soc. London, Ser A {\bf 137}, 696 (1932);
L. Landau, Phys. Z. Sowjetunion {\bf 2}, 46 (1932).

\bibitem{IZ-04}
F. Iachello and N. V. Zamfir, Phys. Rev. Lett. {\bf 92}, 212501 (2004).

\bibitem{LMD-11}
G.-D. Lin, C. Monroe, and L.-M. Duan, Phys. Rev. Lett.
{\bf 106}, 230402 (2011).


\bibitem{GAN-14}
I.M. Georgescu, S. Ashhab, and F. Nori, Rev. Mod.
Phys. {\bf 86}, 153 (2014).

\bibitem{Islam-etal-11}
R. Islam, E.E. Edwards, K. Kim, S. Korenblit,
C. Noh, H. Carmichael, G.-D. Lin, L.-M. Duan,
C.-C. Joseph Wang, J.K. Freericks, and C. Monroe,
Nature Comm. {\bf 2} 377 (2011).

\bibitem{Simon-etal-11}
  J. Simon, W.S. Bakr, R. Ma, M.E. Tai,
P.M. Preiss, and M. Greiner,
Nature {\bf 472}, 307 (2011).

\bibitem{Bloch-08}
I. Bloch, Nature {\bf 453}, 1016 (2008).

\bibitem{YKS-10}
A.P. Young, S. Knysh, and V.N. Smelyanskiy, Phys.
Rev. Lett. {\bf 104}, 020502 (2010).

\bibitem{AC-09}
M.H.S. Amin and V. Choi, Phys. Rev. A {\bf 80},
062326 (2009).

\bibitem{CNPV-15}
M. Campostrini, J. Nespolo, A. Pelissetto, and E. Vicari,
in preparation.



\end{thebibliography}
\end{document}